\documentstyle[12pt]{article}
\begin{document}
\def\thefootnote{\fnsymbol{footnote}}
\begin{flushright}
KANAZAWA-02-14  \\
July, 2002\\  
\end{flushright}
\vspace*{2cm}
\begin{center}
{\LARGE\bf $\mu$-term as the origin of baryon and lepton number asymmetry}\\
\vspace{1 cm}
{\Large  Daijiro Suematsu}
\footnote[1]{e-mail:suematsu@hep.s.kanazawa-u.ac.jp}
\vspace {0.7cm}\\
{\it Institute for Theoretical Physics, Kanazawa University,\\
        Kanazawa 920-1192, JAPAN}
\end{center}
\vspace{2cm}
{\Large\bf Abstract}\\
We study a possibility of combining an origin of the $\mu$-term and
the baryon and lepton number asymmetry. If we assume that the 
$\mu$-term is generated through a flat direction of a singlet scalar
field, the coherent oscillation of this condensate 
around its potential minimum can store the global U(1) charge asymmetry. 
The decay of this condensate can distribute this asymmetry into the 
lepton and baryon number asymmetry as far as its decay occurs at 
an appropriate temperature. 
We examine the compatibility between this scenario and the small 
neutrino mass generation based on both the 
ordinary seesaw mechanism and the bilinear $R$-parity violating terms.
\newpage
\setcounter{footnote}{0}
\def\thefootnote{\arabic{footnote}}
\section{Introduction}
In the present astroparticle physics, it is one of the crucial problems
to clarify an origin of the baryon number $(B)$ asymmetry in the 
universe \cite{bna}.
It has been suggested that the electroweak sphaleron in the standard 
model (SM) plays an important role in that investigation
\cite{spha}. Since the $B$ asymmetry generated at high energy scales 
can be washed out by this effect unless $B-L\not= 0$, we need a 
suitable scenario to escape this problem.
 
Leptogenesis seems to present an elegant scenario for the 
explanation of the $B$ asymmetry on the basis of the lepton number
$(L)$ violation \cite{lept}.
Since the electroweak sphaleron interaction conserves $B-L$, a part of 
the $L$ asymmetry
can be converted into the $B$ asymmetry through this
interaction.
Recent experiments related to the neutrino oscillation suggest 
the existence of small neutrino masses \cite{sol,atm}.
This fact seems to indicate the existence of the $B-L$ violation 
at a certain high energy scale as far as we consider to explain the
small neutrino masses by the seesaw mechanism. 
From this point of view the leptogenesis 
scenario is very promising as an origin of the $B$ asymmetry.

Several leptogenesis scenarios have been proposed under the various 
$L$ violating schemes \cite{lept,lept1}.
In the supersymmetric model, one may consider
the $L$ violation due to some condensates along a flat direction.
As such scenarios there are, for example, 
the decay of the Affleck-Dine (AD) condensate \cite{ad} 
in the $LH_2$ direction \cite{ad1} and
also the decay of the right-handed sneutrino condensate in the chaotic
inflation scenario \cite{sneu,sneu1}.

As an extension of this kind of leptogenesis scenario, we may consider a
possibility that an additional global U(1)$_X$ charge asymmetry 
is generated due to the AD mechanism and then it is
converted into the $B$ and $L$ asymmetry 
through the sphaleron interaction.
This can occur if an interaction which violates both U(1)$_X$
and $B-L$ symmetry is in the thermal equilibrium.\footnote{Similar
scenario is proposed in some works in the different context
\cite{dark,iq,dr}.}   
In this paper, we would like to propose such a model which can
also be related to the origin of the  $\mu$-term 
$\mu H_1H_2$.
 
In the minimal supersymmetric standard model (MSSM)
the origin of a scale of the $\mu$-term is not known.
Since it is a supersymmetric mass term, we have no reason why it
takes a similar value to the supersymmetry (SUSY) breaking scale. 
Fixing its scale to the weak scale is called the $\mu$-problem \cite{mu,mu1}.
Usually the $\mu$-term is considered to be spontaneously generated
through, for example,
a vacuum expectation value (VEV) of some SM singlet scalar field $S$
as $\mu=\lambda\langle S\rangle$.\footnote{ In such a case,
it is well-known that there can be also the Peccei-Quinn symmetry 
as an additional global symmetry.}
If this singlet field has an almost flat direction and 
stores a large energy due to the deviation from its true
vacuum value at a suitable period of the expansion of the 
universe,\footnote{Inflation models based 
on this behavior of $S$ have been considered in \cite{muinf,laz}.
In this paper the inflation is assumed to be induced by other scalar field.}
this condensate may store the sufficient U(1)$_X$ charge asymmetry 
during the oscillation due to the AD mechanism.
The decay of this condensate through the U(1)$_X$ invariant coupling
$\lambda SH_1H_2$ produces the U(1)$_X$ charge asymmetry 
in the Higgsino sector $\tilde H_{1,2} $ and it is distributed into other
fields in the thermal equilibrium.
If both the electroweak sphaleron interaction and a 
certain interaction violating simultaneously both U(1)$_X$ and $B-L$ 
are in the thermal equilibrium at this period,
we expect that a part of this asymmetry is converted into the 
$B$ and $L$ asymmetry.
 
This scenario is largely affected by both temperatures at which the $B-L$ 
violating interaction and the soft SUSY 
breaking effects leave the thermal equilibrium. 
Additionally, the true VEV of $S$ and its coupling $\lambda$ to the Higgsinos 
are constrained by the scale of $\mu$. 
We need to study a viability of this scenario under these constraints.
Moreover, the small mass generation of the neutrinos has an intimate 
connection to this scenario as the origin of the $B-L$ violation, 
although the situation seems to be completely dependent on the model.
For example, one may consider scenarios based on the usual seesaw
mechanism or the bilinear $R$-parity violating terms 
$\epsilon_\alpha L_\alpha H_2$. In the former case, it is well known that 
the lepton number violating effective interaction appears incidentally
at a certain high energy scale. It can leave the thermal equilibrium at a
rather high temperature.
In the latter case, various works suggest 
that the bilinear $R$-parity violating terms can be successfully related 
to the neutrino masses \cite{rparity,hdprv,rparity1}.
However, if the Higgsino-neutrino mixing induced by the bilinear 
$R$-parity violating terms is in the thermal equilibrium around the weak 
scale, the produced $L$ and $B$ asymmetry may be completely washed out.
Thus, it is worthy to study what kind of neutrino mass generation
mechanisms are compatible with this scenario.

This paper is organized as follows. In section 2 we discuss the
production of the global U(1)$_X$ charge asymmetry. 
In section 3 its conversion
into the $L$ and $B$ asymmetry is studied. 
In section 4 we discuss
the relation of this scenario to the neutrino mass. Section 5 is devoted 
to the summary.

\section {Production of the U(1)$_X$ charge asymmetry}
To present the basic idea of the scenario,
we consider a model defined by the superpotential $W_{\rm MSSM}+W_1$.
$W_{\rm MSSM}$ is the superpotential for the MSSM Yukawa 
interactions such as
\begin{equation}
W_{\rm MSSM}=y_U^{\alpha\beta}Q_\alpha\bar U_\beta H_2
+y_D^{\alpha\beta}Q_\alpha\bar D_\beta H_1
+y_E^{\alpha\beta}L_\alpha\bar E_\beta H_1.
\end{equation}
For the construction of an additional superpotential $W_1$ which plays the
essential role in the scenario, we introduce the massless 
SM singlet chiral superfield $S$. As such a superpotential
we suppose
\begin{equation}
W_1=\lambda SH_1H_2 +{d\over M_{\rm pl}^{N-3}}S^N,
\label{eqa}
\end{equation}
where $\lambda$ and $d$ is assumed to be real and $d=O(1)$. 
This superpotential can be realized by imposing a suitable symmetry.
We discuss this point in appendix A.

In this model we find that there are two new abelian global symmetries 
U(1)$_{\rm PQ}$ and U(1)$_R$ other than the $B$ and $L$ symmetries 
if we neglect the nonrenormalizable term in $W_1$. They are defined 
in Table 1. 
Among these four Abelian symmetries there remain two 
global symmetries
U(1)$_{B-L}$ and U(1)$_X$ as those with no SU(3) and SU(2) 
gauge anomaly. It is easily checked that the U(1)$_X$ charge $X$ 
can be represented as the linear combination of the four global 
Abelian charges as $X=B+L-10Q_{\rm PQ}+3Q_R$. 

Now we consider the generation of the U(1)$_X$ charge asymmetry.
For this study we need to note that $S$ gives a $D$-flat direction.
The VEV is generally supposed to be complex. 
The direction described by $\langle S\rangle$ is slightly lifted by 
the nonrenormalizable term in the scalar potential induced by the last 
term of $W_1$. In the early universe there are additional effective 
contributions to the scalar potential induced by the 
SUSY breaking effects caused by 
the large Hubble constant $H$ \cite{dens} and the thermal effects 
\cite{temp} other than the ordinary soft SUSY breaking.
If we take account of these effects and we put 
$\langle S\rangle\equiv ue^{i\theta}$, 
the scalar potential is found to be expressed as 
\begin{eqnarray}
&&V\simeq\left(-cH^2+M_S^2(T)\right)u^2
+{\vert d\vert^2N^2\over M_{\rm pl}^{2N-6}}u^{2N-2} \nonumber\\
&&\hspace*{2cm}+\left\{\left({am_{3/2}e^{i\theta_a}\over M_{\rm pl}^{N-3}}
+{bHe^{i\theta_b} \over M_{\rm pl}^{N-3}}\right)
u^Ne^{iN\theta}+{\rm h.c.}\right\},
\label{eqaa}
\end{eqnarray} 
where $m_{3/2}$ is a typical soft SUSY breaking scale of $O(1)$~TeV 
and the coefficients $a$, $b$ and $c$ are $O(1)$ real constants. 
The CP phases $\theta_a$ and $\theta_b$ in the curly brackets 
are induced by the above mentioned SUSY breaking effects 
which violate the U(1)$_X$.  
The effective mass $M_S^2(T)$ contains the usual soft SUSY 
breaking mass $m_S^2$ of $O(m_{3/2}^2)$ and the thermal mass 
$\sim\lambda^2T^2$ caused by the 
coupling of $S$ with $H_{1,2}$ in the thermal plasma. 
It can be expressed depending on the value of $u$ as \cite{temp}
\begin{equation}
M_S^2(T)\simeq\left\{\begin{array}{ll}m^2_S & \quad (\lambda u>T), \\
m^2_S+\lambda^2 T^2 & \quad (\lambda u<T). \\
\end{array}\right.
\label{therm}
\end{equation} 

\begin{figure}[tb]
\small
\begin{center}
\begin{tabular}{c|cccccccccccc}
  &$\tilde g$, $\tilde W$, $\tilde B$  & $Q_L$ &$\bar U_L$ &$\bar D_L$ & $L_L$ 
 &$\bar E_L$ &$H_1$ &$H_2$ & $S$
 & $A_{SU(3)}$& $A_{SU(2)}$   \\ \hline
$Q_{\rm PQ}$ &  0 &  0 &$-2$ &1 & $-1$ &2 &$-1$ & 2 &  $-1$ &
  $-{3\over 2}$ & $-1$\\
$Q_R$  &$-1$ & $-{1\over 6}$ &$-1$ & 0 &$-{5\over 6}$ 
&${2\over 3}$ &$-{5\over 6}$ & ${1\over 6}$ &$-{1\over 3}$ &
$-5$ & $-{13\over 3}$\\
$B$  &  0 & ${1\over 3}$  &$-{1\over 3}$ & $-{1\over 3}$ & 
0 & 0 & 0 &0 & 0  &0 & ${3\over 2}$\\
$L$    & 0    & 0   & 0   & 0  & $1$   & $-1$ & 0   & 0 & 
0  & 0 & ${3\over 2}$   \\ \hline
\end{tabular}
\vspace*{3mm}\\
Table 1~ Global U(1) charge assignment and gauge 
anomaly.
\end{center}
\end{figure}

\normalsize
The Hubble constant contribution $H^2$ dominates the mass of the condensate
during the inflation.
Thus, if the sign of the Hubble constant contribution in
eq.~(\ref{eqaa}) is negative $(c>0) $\cite{dens}, 
the magnitude of the condensate takes a large value such as
\begin{equation}
u_I\simeq \left(H M_{\rm pl}^{N-3}\right)^{1\over N-2}.
\label{eqa3}
\end{equation} 
On the other hand,
the phase $\theta$ of the condensate at the potential minimum takes
 one of the $N$ distinct values 
$\theta=-\theta_b/N+2\pi \ell/N~(\ell=1, 2, \cdots, N)$.
At this period the condensate follows this instantaneous 
potential minimum since its evolution is almost the critical damping.
The dilute plasma appears as a result of a partial decay of the
inflaton. Then the temperature rapidly increases to 
$T_{\rm max}\simeq (T_R^2H M_{\rm pl})^{1/4}$ \cite{temp}.
$T_R$ is the reheating temperature realized after the completion 
of the inflaton decay and can be expressed as 
$T_R\simeq\sqrt{M_{\rm pl}\Gamma_I}$ where $\Gamma_I$ is the inflaton
decay width.
If this temperature $T_{\rm max}$ does not satisfy 
$\lambda\vert u_I\vert<T_{\rm max}$, 
no thermal contribution to $M_S^2(T)$ 
appears and $M_S^2(T)$ takes the expression of the upper one 
in eq.~(\ref{therm}) \cite{dens,temp}. 
Thus, the condition for the thermal effects to be negligible during the
inflation gives the following lower bound on $\lambda$:
\begin{equation}
\lambda > T_R^{1/2}H_I^{N-6\over 4(N-2)}M_{\rm pl}^{10-3N\over 4(N-2)},
\label{eqa4}
\end{equation}
where $H_I$ is the Hubble parameter during the inflation.

When the inflaton evolves and $H$ decreases to $H\sim m_{3/2}$,  
the effective squared mass of the condensate becomes positive and 
then $u=0$ is the minimum of the scalar potential $V$.
 The condensate starts to oscillate around $u=0$, and the
thermal effects due to the dilute plasma to $M_S^2(T)$ appears. 
Then $M_S^2(T)$ takes the form of the lower one 
in eq.~(\ref{therm}). At this time, the dominant term for the 
U(1)$_X$ breaking changes from the second term in the curly brackets 
of eq.~(\ref{eqaa}) to the first term. 
Since the phase $\theta_a$ and $\theta_b$ are generally independent,
the phase $\theta$ of the condensate changes non-adiabatically from 
that determined by $\theta_b$ to that determined by $\theta_a$ 
due to the torque in the angular direction. 
Thus, the $X$ asymmetry is stored in the condensate during its 
evolution due to the AD mechanism \cite{ad}.

This $X$ asymmetry can be estimated by taking account that 
the $X$ current conservation is violated by the dominant $X$ breaking
term in the curly brackets in eq.~(\ref{eqaa}) as
\begin{equation}
{d\delta n_X(t)\over dt}= X(S){am_{3/2}u^N\over M_{\rm pl}^{N-3}}\sin\delta,
\end{equation} 
where $X(S)$ is the $X$ charge of $S$ and $\delta$ is determined 
by the difference of $\theta_a$ and $\theta_b$. 
By solving this equation, the $X$ asymmetry produced at this period 
is found to be roughly expressed as 
\cite{ad,dens,temp}\footnote{The rigorous estimation requires the
numerical calculation as discussed in \cite{dens}. 
It is beyond the scope of this paper and we do not go further here.}
\begin{equation}
\delta n_X(t)\simeq X(S){m_{3/2}\over H}H^{N\over N-2}
M_{\rm pl}^{2(N-3)\over N-2}\sin\delta,
\end{equation}
where $t$ is the time when $H\sim m_{3/2}$.  Following this, the
reheating due to the inflaton decay is completed at 
$H\sim \Gamma_I$.\footnote{Here we assume that $m_{3/2}>\Gamma_I$ is
satisfied. This means $T_R<\sqrt{M_{\rm pl}m_{3/2}}\simeq 
10^{11}{\rm GeV}$.}
On the other hand, the stored $X$ asymmetry in the condensate is 
liberated into the thermal plasma through the decay of the condensate 
by the $X$ conserving coupling $\lambda SH_1H_2$ in $W_1$.
Since the oscillation behaves as a matter for the expansion of the 
universe, it can dominate
the energy density of the universe before its decay which occurs at 
$H\sim \Gamma_S$. For the reasonable value of $\lambda$, this is the
case since $\Gamma_S<\Gamma_I$ is satisfied. 
Taking account of this, we estimate the ratio of the $X$ asymmetry to 
the entropy density $s$ at this period as
\begin{equation} 
{\delta n_X(\tilde t_R)\over s}
\simeq {\delta n_X(t)\over \tilde T_R^3}{t^2\over \tilde t_R^2}
\simeq X(S)\tilde T_Rm_{3/2}^{4-N\over N-2}M_{\rm pl}^{-2\over N-2}
\sin\delta, 
\label{eqe}
\end{equation}
where we use $\tilde t_R\sim \Gamma_S^{-1}\sim M_{\rm pl}/\tilde T_R^2$. 
If the temperature $\tilde T_R(\simeq 10^{10}\lambda~{\rm GeV})$ 
is appropriate to keep the $X$ asymmetry and
convert it into the $B$ asymmetry through the sphaleron interaction, 
we can obtain the $B$ asymmetry.\footnote{This $\tilde T_R$ is the
marginal value for the cosmological gravitino problem \cite{grav}.}  

Finally we discuss the evolution of $u$ after this period to
see the relation of this scenario to the $\mu$-term.
We assume $H_I\sim 10^{13}$~GeV during the inflation on the basis of the CMB 
data and $T_R~{^<_\sim}~10^9$~GeV. For these values we obtain 
$T_{\rm max}\sim 10^{13}$~GeV. Thus, if we take $N=4$ in $W_1$, 
as an example, eq.~(\ref{eqa3}) gives $u_I\sim 10^{16}$~GeV and  
eq.~(\ref{eqa4}) suggests that $\lambda~{^>_\sim}~ 10^{-3.5}$ should 
be satisfied. 
When the temperature decreases from $T_{\rm max}$ to 
$T_c\sim m_{3/2}/\lambda$, $M_S^2(T)$ represented by the
lower one in eq.~(\ref{therm}) starts to be dominated by the
 soft supersymmetry breaking mass $m_S^2$.
If $m_S^2<0$ is realized by some reason \cite{rad},
$u\not= 0$ becomes the true vacuum after this period. 
Since the $\mu$-term is generated from the first term in 
$W_1$ as $\mu=\lambda u$,  
such a value of $u$ should be $u_0~{^<_\sim}~10^6$~GeV to realize the
appropriate $\mu$ for the above mentioned $\lambda$.\footnote{Such a
$u_0$ may be expected to be determined either by the nonrenormalizable 
terms or by the pure radiative symmetry breaking effect 
in which $u_0$ is estimated by using the renormalization group equation 
as discussed in \cite{rad1}.}
Although the condensate again starts to oscillate around $u_0$, 
the deviation from $u_0$ instantaneously decays into the light fields 
through the $X$ conserving coupling $SH_1H_2$ 
since $H<\Gamma_S$ is satisfied at this time.  
The released energy cannot dominate the total energy density
$({\pi^2\over 30}g_\ast T^4 \gg m_{3/2}^2u_0^2)$ and then the 
effects of the produced entropy is negligible.
The $X$ asymmetry obtained in eq.~(\ref{eqe}) can be used as 
the origin of the $B$ asymmetry. 

\section{Generation of $B$ asymmetry}
Next we examine whether this $X$ asymmetry transformed into
the thermal plasma through the decay of the condensate can remain 
as a nonzero value and be partially converted into the $B$ asymmetry. 
This should be studied taking account that the electroweak 
sphaleron interaction and other various interactions are in 
the thermal equilibrium.
For this study, it is convenient to consider the detailed balance of
these interactions and solve
the chemical equilibrium equations \cite{lb,bp}. The particle-antiparticle
number asymmetry $\delta n_f$ can be approximately related 
to the corresponding 
chemical potential $\mu_f$. In the case of $\mu_f \ll T$, it can be
represented as
\begin{equation}
\delta n_f\equiv n_{f} -n_{f^c}=\left\{ \begin{array}{ll}
\displaystyle
{g_f\over 6}T^2\mu_f & (f~:~{\rm fermion}),\\
\displaystyle
{g_f\over 3}T^2\mu_f & (f~:~{\rm boson}), \\ \end{array} \right. 
\label{eqee}
\end{equation}
where $g_f$ is a number of relevant internal degrees of 
freedom of the field $f$.
By solving the detailed valance equations for the chemical potential $\mu_f$,
we can study the charge asymmetry at the period 
after the decay of the condensate.

If the SU(2) and SU(3) sphaleron interactions are in the thermal
equilibrium, we have the conditions such as
\begin{eqnarray}
&&\sum_{i=1}^{N_g}\left(3\mu_{Q_i}+\mu_{L_i}\right)+\mu_{\tilde{H_1}}
+\mu_{\tilde{H_2}}+ 4\mu_{\tilde W}=0,
\label{eqee1} \\  
&&\sum_{i=1}^{N_g}\left(2\mu_{Q_i}-\mu_{U_i}-\mu_{D_i}\right)
+6\mu_{\tilde g}=0,
\end{eqnarray}
where $N_g$ is a number of the generation of quarks and leptons.
The cancellation of the total hypercharge or the electric charge 
of plasma in the universe requires
\begin{eqnarray}
&&\sum_{i=1}^{N_g}\left(\mu_{Q_i}+2\mu_{U_i}-\mu_{D_i}-\mu_{L_i}-
\mu_{E_i}\right)+\mu_{\tilde{H_2}}-\mu_{\tilde{H_1}} \nonumber \\
&&\hspace{2cm}+2\sum_{i=1}^N(\mu_{\tilde{Q_i}}+2\mu_{\tilde{U_i}}-
\mu_{\tilde{D_i}}-\mu_{\tilde{L_i}}-\mu_{\tilde{E_i}})
+2\left(\mu_{H_2}-\mu_{H_1} \right)=0.
\end{eqnarray}
When Yukawa interactions in $W_{\rm MSSM}+W_1$ are in the thermal 
equilibrium, they impose the conditions\footnote{
We should note that the second term in $W_1$ leaves the thermal equilibrium 
at $T\sim M_{\rm pl}$. Since $S$ has no other coupling to the MSSM
contents than $\lambda SH_1H_2$, the last one in eq.~(\ref{chemi}) is
the only condition for $\mu_S$.}
\begin{eqnarray} 
&&\mu_{Q_i}-\mu_{U_j}+\mu_{H_2}=0, \quad
\mu_{Q_i}-\mu_{D_j}+\mu_{H_1}=0, \nonumber \\
&&\mu_{L_i}-\mu_{E_j}+\mu_{H_1}=0, \quad
\mu_S +\mu_{\tilde H_1}+\mu_{\tilde H_2}=0.
\label{chemi}
\end{eqnarray}
There are also the conditions for the gauge interactions in the
thermal equilibrium, which are summarized as
\begin{equation}
\mu_{\tilde{Q_i}}=\mu_{\tilde g}+\mu_{Q_i}=\mu_{\tilde W}+\mu_{Q_i}
=\mu_{\tilde B}+\mu_{Q_i}.
\label{eqee2}
\end{equation}
The similar relations to eq.~(\ref{eqee2}) is satisfied 
for leptons $L_i$, Higgs $H_{1,2}$ and other
fields $U_i, D_i,  E_i $ which have the SM gauge interactions.
Flavor mixings of quarks and leptons due to the Yukawa couplings
allow us to consider the flavor independent chemical potential such as
$\mu_{Q}=\mu_{Q_i}$ and $\mu_{L}=\mu_{L_i}$.

Here we introduce a term violating both $B-L$ and $X$, which is
necessary to convert the $X$ asymmetry into the $B$ and $L$ asymmetry.
If such a term exists, only a linear combination of these two U(1)s is 
absolutely conserved. Then a part of $X$ asymmetry can be converted
into the $B-L$ asymmetry.
As an interesting example, we may consider a term $(LH_2)^k$.
It corresponds to the effective neutrino mass term of the ordinary
seesaw mechanism in the $k=2$ case and also the bilinear $R$-parity 
violating term in the $k=1$ case.
The thermal equilibrium condition of these terms can be written as 
\begin{equation}
\mu_L+\mu_{H_2}=0 \quad (k=2), \qquad \mu_L+\mu_{\tilde H_2}=0 \quad (k=1).
\label{vlep}
\end{equation}
We find that there is an independent chemical potential in these 
thermal equilibrium conditions (\ref{eqee1})$\sim$(\ref{vlep}).
It can be taken as $\mu_{\tilde H_2}$, which corresponds to 
the above mentioned remaining symmetry.

By now we have not taken account of the equilibrium 
conditions for the soft SUSY breaking terms. 
The soft SUSY breaking terms are in the thermal equilibrium 
when $H~{^<_\sim}~\Gamma_{ss}$ is satisfied. 
Here the rate of the soft SUSY breaking effects is written as 
$\Gamma_{ss}\simeq m_{3/2}^2/T$ \cite{iq}. 
From this we find that the soft SUSY breaking effects are 
in the thermal equilibrium for the temperature 
$T~{^<_\sim}~T_{ss}\simeq 10^7$~GeV. 
Thus, for $T~{^<_\sim}~T_{ss}$ we find that $\mu_{\tilde g}=0$ is satisfied 
and then eqs.~(\ref{eqee1}) $\sim$ (\ref{vlep}) result 
in $\mu_{\tilde H_2}=0$. 
The $X$ asymmetry produced through the decay of the condensate
disappears in this case. In order to escape this, 
if we define $T_X$ as a temperature
at which the $X$ and $B-L$ violating interaction is out-of-equilibrium,
we need to require that $T_X$ should satisfy $T_{ss}~{^<_\sim}~T_X
~{^<_\sim}~\tilde T_R$. We will discuss this condition in the next section.
If these conditions are satisfied, the $X$ asymmetry induced through 
the decay of the condensate can be partially converted into 
the $B$ asymmetry.

By solving eqs.~(\ref{eqee1}) $\sim$ (\ref{vlep}),
$\mu_Q$, $\mu_L$, $\mu_{H_{1,2}}$ and $\mu_{\tilde g}$ can be written 
with the chemical potential of Higgsino field $\tilde H_2$ 
at $T_X$ in such a way as
\begin{eqnarray}
&&\mu_Q={17N_g+6\over N_g(10N^2_g-17N_g-15)}\mu_{\tilde H_2}, \quad
\mu_L=-\mu_{H_2}={5(4N_g+3)\over 10N^2_g-17N_g-15}\mu_{\tilde H_2}, 
\nonumber \\
&&\mu_{H_1}=-{40N_g+3\over 10N^2_g-17N_g-15}\mu_{\tilde H_2}, \quad
\mu_{\tilde g}=-{(10N_g+3)N_g\over 10N^2_g-17N_g-15}\mu_{\tilde H_2},
\label{eqee3}
\end{eqnarray}
for the $k=2$ case and
\begin{eqnarray}
&&\mu_Q={2N^3_g+N_g^2+17N_g+6\over N_g(6N_g^2-17N_g-15) }
\mu_{\tilde H_2},\quad
\mu_L=-\mu_{\tilde H_2}, \quad
\mu_{H_2}=-{4N^2_g+20N_g+15\over 6N_g^2-17N_g-15}\mu_{\tilde H_2},
\nonumber \\
&&\mu_{H_1}={4N^2_g-40N_g-3\over 6N_g^2-17N_g-15}\mu_{\tilde H_2},\quad
\mu_{\tilde g}=-{(10N_g+3)N_g\over 6N_g^2-17N_g-15}\mu_{\tilde H_2},
\label{eqee4}
\end{eqnarray}
for the $k=1$ case.
Defining $B$ and $L$ as $\delta n_B\equiv BT^2/6$ 
and $\delta n_L\equiv LT^2/6$, 
we can calculate these values at $T_{ss}$ by using eq.~(\ref{eqee}), 
(\ref{eqee3}) and (\ref{eqee4}) as  
\begin{eqnarray}
&&k=2:~ \left\{\begin{array}{l}
\displaystyle B={80N_g^3+204N_g^2-150N_g-72\over 360N_g^3
+3308N^2_g-1419N_g-1143}X, \\  
\displaystyle L={N_g(60N^2_g-42N_g-126)\over 360N_g^3
+3308N^2_g-1419N_g-1143 }X, \\
\end{array}\right. 
\label{eqeed} \\
&&k=1:~
\left\{\begin{array}{l}
\displaystyle B={56N^3_g+192N^2_g-150N_g-72\over 378N_g^3
+3822N_g^2-624N_g-1143}X, \\
\displaystyle L={N_g(92N^2_g-15N_g-126)\over 378N_g^3
+3822N_g^2-624N_g-1143}X,\\ 
\end{array}\right.
\label{eqeee}
\end{eqnarray}
where $X$ is defined as $\delta n_X\equiv XT^2/6$.
These results show that all of $B$, $L$ and $B-L$ take nonzero 
values as far as $X\not= 0$.

When the temperature goes below $T_{ss}$, 
the soft SUSY breaking terms are in the thermal equilibrium.
This results in $\mu_{\tilde g}=0$ and $X=0$.
However, if the $X$ and $B-L$ violating interaction is assumed to 
be out-of-equilibrium 
at $T_{ss}$, the equilibrium conditions are represented by
(\ref{eqee1})$\sim$(\ref{eqee2}). Thus the $B-L$ asymmetry existing at
$T_{ss}$ is kept after this period. The equilibrium conditions 
give the ordinary MSSM values for $B$ and $L$ as
\begin{equation}
B={4(2N_g+1)\over 22N_g+13}(B-L), \qquad L=-{14N_g+9\over 22N_g+13}(B-L),
\end{equation}
where we should use the $B-L$ value obtained from 
eq.~(\ref{eqeed}) or (\ref{eqeee}).
The $B$ asymmetry produced in this scenario is finally
estimated as
\begin{equation}
Y_B\equiv {\delta n_B\over s} ={\delta n_X\over s}~{B-L\over X}
~{4(2N_g+1)\over 22N_g+13}~\kappa 
\simeq \tilde T_R~m_{3/2}^{4-N\over N-2}M_{\rm pl}^{-2\over N-2}~f(N_g)
~\kappa\sin\delta,
\label{eqe4}
\end{equation}
where $\kappa (\le 1)$ is introduced to take account of the washout 
effect which we will discuss later.
$f(N_g)$ is a numerical factor taking $f(3)\simeq 0.3$ and 0.05 
for $k=2$ and $k=1$, respectively.
From this result, we find that this scenario can produce the presently
observed $B$ asymmetry $Y_B=(0.6~-~1)\times 10^{-10}$ as far as $N\ge 5$
for $\tilde T_R>10^7~{\rm GeV}$.
Since the $X$ asymmetry disappears at the temperature 
less than $T_{ss}$ through the effects of the soft SUSY breakings, 
the $X$ asymmetry cannot be related to the dark matter
abundance in the universe \cite{dark}. 

\section{Relation to neutrino masses}
It is an interesting and important issue to consider the compatibility 
of this scenario for the generation of the $B$ and $L$ asymmetry 
with the small neutrino mass generation.
This is because the $X$ and $B-L$ violation is considered to be 
introduced through the interaction related to the neutrino masses as
mentioned before.
In this section we study two typical schemes for the neutrino mass 
generation in this scenario.
For this purpose we introduce SM singlet chiral superfields $\bar{\cal N}$ 
and ${\cal N}$ which have the global U(1) charges listed in Table.~2.

\begin{figure}[tb]
\begin{center}
(a) \quad\begin{tabular}{c|cc}
  & $Q_{\rm PQ}$ & $Q_R$ \\ \hline
$\bar{\cal N}$& 0 & $-{1\over 2}$ \\
\end{tabular}
\hspace*{2cm}
(b)\quad\begin{tabular}{c|cc}
  & $Q_{\rm PQ}$ & $Q_R$ \\ \hline  
$\bar{\cal N}$& $-{1\over 3}$ & $-{3\over 4}$ \\
${\cal N}$     & ${1\over 2}$  & ${1\over 4}$  
\end{tabular}
\vspace*{3mm}\\
{\footnotesize Table~2~~The U(1) charge assignments 
for $\bar{\cal N}$ and ${\cal N}$, whose $B$ and $L$ are zero.}
\end{center}
\end{figure}

\subsection{Ordinary see-saw scenario}
First we consider the lepton sector characterized by the 
renormalizable superpotential
\begin{equation}
W_2=y_N^{\alpha}L_\alpha\bar{\cal N} H_2 
+ M_R\bar{\cal N}^2 +\cdots.
\end{equation}
Here we assume the (a) type charge assignment in Table 2 and abbreviate the
irrelevant terms.\footnote{There is another possible term $\bar{\cal
N}^3$, which violates U(1)$_R$ and is the same order as the presented one.
However, it brings no influence for the present argument.} 
The first term violates both U(1)$_{B-L}$ and U(1)$_X$.
If we integrate out the right-handed neutrino $\bar{\cal N}$ following
the usual seesaw scheme, we have the effective $X$ violating 
interaction which corresponds to the $k=2$ case
discussed in the previous section.
The necessary condition for the applicability of this scenario is that
this effective interaction should have left the 
thermal equilibrium by $T_{ss}$.
By using the right-handed neutrino mass $M_R$,
we can summarize this condition such that 
$H>T_X^3/M^2_R$ should be satisfied at 
$T_X~{^>_\sim}~T_{ss}$. 
This results in $M_R~{^>_\sim}~10^{12}$~GeV which is 
a suitable value for the explanation of the neutrino masses required by
the neutrino oscillation data \cite{sol,atm}.
 
In this type of seesaw model the leptogenesis is usually considered 
on the basis of the out-of-equilibrium decay of the heavy right-handed 
neutrinos or the decay of sneutrino condensate. 
However, if we consider the spontaneous $\mu$-term generation
along the almost flat direction of $\langle S\rangle$ as discussed here, 
the $B$ asymmetry produced by this usual leptogenesis 
might not be the dominant one.
As mentioned before, we assume that the decay of the condensate
is completed above the temperature $T_X$ which can be sufficiently lower
than the masses of the right-handed heavy neutrinos.
Then the $B$ asymmetry produced through the usual 
scenario seems to be washed out or overridden by the $B$ asymmetry 
produced in the present scenario.
If there is an additional entropy production after the thermal
decoupling of the weak sphaleron interaction, such a reheating 
temperature $T_N$ should satisfy $T_N~{^>_\sim}~1$~MeV because of the
nucleosynthesis requirement. Since the dilution effect $\kappa$ due to 
this entropy release is written as $\kappa\sim T_N/T_W$ where $T_W\simeq
100$~GeV, we can estimate it as $\kappa~{^>_\sim}~10^{-5}$.
Thus the baryon number asymmetry produced by the present scenario 
with an appropriate value of $N$ can be sufficient for the 
explanation of the baryon number in the universe.
This kind of possibility may be promising as much as 
the usual leptogenesis scenario.

\subsection{Bilinear $R$-parity violating scenario}
Another scheme for the small neutrino mass may be characterized by 
the bilinear $R$-parity violating term in the superpotential
such as
\begin{equation}
W_3=\epsilon_\alpha L_\alpha H_2,
\end{equation} 
which violates both U(1)$_{B-L}$ and U(1)$_X$.
In the MSSM with $W_3$ we can show that the small neutrino mass 
generation is possible due to the neutralino-neutrino mixing.
We discuss this point in the appendix B.

If bilinear $R$-parity violating term is
simultaneously in the thermal equilibrium with the soft SUSY
breaking terms, the chemical detailed balance equations 
result in $\mu_{\tilde H_2}=0$. 
In order to keep nonzero $B-L$ asymmetry,
the $\epsilon_\alpha L_\alpha H_2$ term should
be in the thermal equilibrium at $T~{^>_\sim}~T_{ss}$
but be out-of-equilibrium at $T<T_{ss}$. 
We can show that such a required behavior 
of the bilinear $R$-parity violating term is realizable by presenting 
a simple example.
We assume the type (b) charge assignment for the SM singlet
chiral superfields ${\cal N}$ and $\bar{\cal N}$.
If we impose either U(1)$_{\rm PQ}$ or U(1)$_R$ invariance for the
MSSM contents, we have the superpotential
\begin{equation}
W_4=\hat c_1{\bar{\cal N}^3\over M^2}LH_2 
+ \hat c_2{(\bar{\cal N}{\cal N})^2 \over M}+
+\hat c_3{\bar{\cal N}^3{\cal N}^2 \over M^2}+\cdots,
\label{eqf1}
\end{equation} 
where $\hat c_{1,2,3}$ are $O(1)$ constants and the ellipses 
represent higher order terms.
The first term violates both U(1)$_{B-L}$ and U(1)$_X$.
If we introduce the soft SUSY breaking masses for ${\cal N}$ and 
$\bar{\cal N}$ whose square is assumed to be negative 
and $O(m_{3/2}^2)$, the scalar 
components of ${\cal N}$ and $\bar{\cal N}$ obtain the nonzero VEV such as 
$\langle {\cal N}\rangle=\langle \bar{\cal N}\rangle\sim\sqrt{m_{3/2}M}$. 
The bilinear $R$-parity violating term in $W_3$ is generated by this VEV
from the first term in eq.~(\ref{eqf1}) as 
$\epsilon=\langle\bar{\cal N}\rangle^3/M^2$.\footnote{
If $M$ is assumed to be an intermediate mass scale such as 
$M~{^>_\sim}~ 5\times 10^9$~GeV,
we find that $\langle\bar{\cal N}\rangle~{^>_\sim}~ 3\times 10^5$~GeV and 
$\epsilon~{^<_\sim}~ 7\times 10^{-4}$~GeV. 
These values suggest that the neutrino masses based on the bilinear 
$R$-parity violation can be the dominant one for the neutrino masses 
in comparison with the usual seesaw contribution.}

The important problem is at what temperature the first term 
in eq.~(\ref{eqf1}) leaves the thermal equilibrium.
If the first term is assumed to leave the thermal equilibrium at $T_X$,
the third term is also out of equilibrium 
but the second term still remains in the thermal equilibrium at $T_X$. 
This means that both $\mu_{\cal N}=\mu_{\bar{\cal N}}=0$ and 
$\mu_L+\mu_{\tilde H_2}=0$ can be satisfied at $T~{^>_\sim}~T_X$.
This result is found to realize the same one which is 
given in eq.~(\ref{vlep}).

On the other hand, the effectively induced $W_3$ should be 
out of equilibrium at $T~{^>_\sim}~T_W$
since it violates the $L$ invariance.\footnote{This condition 
may be expressed that $H>\epsilon_\alpha^2/T$ is satisfied at 
$T~{^>_\sim}~T_W(\simeq 100~{\rm GeV})$.}
This gives the constraint on $\epsilon_\alpha$.
It is convenient to redefine the chiral superfield $H_1$ as 
$H_1^\prime\equiv {\epsilon_\alpha\over \mu}L_\alpha +H_1$ to estimate it.
By this manipulation, $\epsilon_\alpha L_\alpha H_2$ disappears from the
superpotential $W_4$ but there appear the new $R$-parity violating terms
\begin{equation}
W_{\rm RPV}=
-y_E^{\alpha\beta}{\epsilon_\gamma\over\mu}L_\alpha\bar E_\beta L_\gamma
-y_D^{\alpha\beta}{\epsilon_\gamma\over\mu}Q_\alpha\bar D_\beta L_\gamma.
\label{eqff}
\end{equation}
These interactions should not completely wash out the $B$ and 
$L$ asymmetries by the collaboration with the sphaleron interaction.
The stringent constraint on these terms is derived by requiring 
that the $L$ violating scattering processes are out of thermal 
equilibrium at $T~{^>_\sim}~T_W$. 
By estimating these processes at $T_W$, we can find a condition 
such as \cite{dr,bp}
\begin{equation}
\left\vert~ y_{E,D}^{\alpha\beta}~{\epsilon_\gamma\over\mu}~ 
\right\vert~{^<_\sim}~10^{-7}. 
\label{eqfff2}
\end{equation} 
Since these Yukawa coupling constants $y_{E,D}$ are constrained by the
masses of quarks and leptons, eq.~(\ref{eqfff2}) 
gives us a condition on $\vert\epsilon_\alpha/\mu\vert$.
If we assume 
$\langle H_1^\prime\rangle=100$~GeV, we find that
$\vert\epsilon_\alpha/\mu\vert~{^<_\sim}~10^{-6}$.\footnote{
This constraint is much stronger than the ones obtained from
the accelerator experiments \cite{rconst}.
The thermal corrections on the Higgsino masses 
is not expected to be so large as to make this bound 
change substantially.}  
Since the constraint is very critical for this neutrino mass generation 
scheme in the present scenario, the more quantitative study 
seems to be necessary.
In fact, in order to obtain the appropriate $B$ asymmetry 
in eq.~(\ref{eqe4}), 
the substantial dilution $\kappa$ seems to be necessary for $N\ge 5$
as far as we do not require the extreme fine tuning of $\delta$.

The washout factor $\kappa$ can be quantitatively estimated by the analysis 
of the Boltzmann equation for the particle 
number density, which includes the effects of the $L$ violating two body 
scatterings. 
It can be written as \cite{bna,bp}
\begin{equation}
{d N_L\over dx}=-{\Gamma_A\over H(m)}{x\over N_L^{\rm EQ}}\left(N_LN_a-
N_L^{\rm EQ}N_a^{\rm EQ}\right),
\label{bol}
\end{equation}
where $N_f\equiv n_f/s$ stands for the number density of $f$ per comoving 
volume and $N_f^{\rm EQ}$ represents the value at the thermal equilibrium.
A dimensionless parameter $x=m/T$ is used for a certain mass scale $m$. 
$H(m)$ is the Hubble constant at $T=m$.
Eq.~(\ref{bol}) can be deformed into the equation for 
the $L$ asymmetry
$Y_L\equiv N_L- N_{L^c}$.  From this equation we can obtain the 
washout factor $\kappa$ as
\begin{equation}
\kappa= \exp\left(-\int_1^{x_0} dx{x\Gamma_A\over H(m)}
{Y_a^{\rm EQ}\over Y_L^{\rm EQ}}\right),
\label{eqf4}
\end{equation}
where $x_0$ should be taken as a value at which the electroweak sphaleron
interaction can be regarded to be out of equilibrium. 
If we assume that $m$ is the SUSY breaking scale $O(1)$~TeV
which seems to be a reasonable choice in the present scenario,
$x_0$ should be fixed as 10. 

In the calculation of $\kappa$, the scattering process $LQ
\rightarrow \lambda_3\bar D$ and its SUSY 
partner become the dominant contribution to the reaction rate 
$\Gamma_A$ in eq.~(\ref{eqf4}).
For these tree diagrams, it can be, respectively, represented as
\begin{equation}
\Gamma_A\sim\alpha_g\alpha_{y_{E,D}}\left(\epsilon\over\mu\right)^2
{T^5\over (T^2 +m^2)^2},
\qquad
\Gamma_A\sim\alpha_g\alpha_{y_{E,D}}\left(\epsilon\over\mu\right)^2T,
\end{equation}
where we use $\alpha_g=g^2/4\pi$ and so on.
For simplicity, we take the masses of all superpartners as $m$. 
Using these $\Gamma_A$'s and practicing the numerical calculation of
eq.~(\ref{eqf4}), we find that $\kappa$ takes the values 
shown in Table~3 for various values of $\vert\epsilon/\mu\vert$. 
We also find that the value of $\kappa$ becomes constant at 
$x_0~{^>_\sim}~10$ for these values of $\vert\epsilon/\mu\vert$. 
It means that these $L$ violating
processes completely leave the thermal equilibrium at $T\sim T_W$.
For $\vert\epsilon/\mu\vert > 10^{-5}$, the value 
of $\kappa$ is less than $10^{-11}$ and we cannot obtain the sufficient 
$B$ asymmetry in any case.
\begin{figure}[tb]
\begin{center}
\begin{tabular}{cc|cc}
$\vert\epsilon/\mu\vert$ & $\kappa$&~~ 
$\vert\epsilon/\mu\vert$ 
& $\kappa$\\ \hline
$10^{-6}$ & 0.85 & $10^{-5.2}$ & $1.5\times 10^{-3}$ \\
$10^{-5.5}$ & 0.2 & $10^{-5.1}$ & $3.5\times 10^{-5}$  \\ 
$10^{-5.3}$ & $1.7\times 10^{-2}$ & $10^{-5}$ & $8.6\times 10^{-8}$ \\ \hline
\end{tabular}
\vspace*{3mm}

{\footnotesize Table~3~~The washout effect $\kappa$ in the bilinear $R$-parity
 violating model.}
\end{center}
\end{figure}
This estimation of $\kappa$ shows that 
the magnitude of the $R$-parity violating coupling is severely
constrained by the washout effect. 
However, the neutrino mass generation scheme based on the bilinear $R$-parity
violation seems to be compatible to the present scenario. 
This washout effect may make it possible to produce 
the appropriate $B$ asymmetry without any fine tuning of the value of
$CP$ phases for a certain range of $\vert\epsilon/\mu\vert$.

\section{Summary}
We have proposed the baryogenesis scenario 
which is intimately related to the origin of the $\mu$-term and the small
neutrino masses. 
If the $\mu$-term is assumed to be originated from a suitable
VEV of the SM singlet field $S$ with the flat direction, 
the deviation of this condensate 
from the true vacuum value induces the coherent oscillation.
If its nonrenormalizable interaction violates the global U(1)$_X$ charge at
the early stage of the universe and there is the non-adiabatic change of
the $CP$ phase during the oscillation, the condensate can store 
the $X$ asymmetry due to the AD mechanism.  
Since its decay proceeds through the U(1)$_X$ invariant interaction, this
asymmetry is distributed into the thermal plasma. 
If a suitable interaction which violates both U(1)$_{B-L}$ and 
U(1)$_X$ is in the thermal equilibrium at the
temperature $T_X$ where the decay of condensate has been almost completed, 
this $X$ asymmetry can be converted into the $L$ and $B$ asymmetry.
  
The required condition for the success of 
this scenario is that the temperature $T_X$ should be higher than
$T_{ss}$ at which the soft SUSY breaking terms come in the thermal
equilibrium. Moreover, the interaction violating both U(1)$_{B-L}$ and 
U(1)$_X$ should leave the thermal equilibrium at the temperature $T$
such as $T_{ss}~{^<_\sim}~T~{^<_\sim}~T_X$. 
If these conditions are satisfied,   
the conversion of the $X$ asymmetry into the $B$ 
asymmetry can be substantially proceeded.
The produced abundance of the $B$ asymmetry 
can be related to the dimension of the U(1)$_X$ violating interaction.
We have also studied the compatibility with two types of $L$ 
violating interactions which are intimately related to the small 
neutrino mass generation.

We have left some problems as those beyond the scope of this paper.
One of them is to present the concrete models which cause the final
stage symmetry breaking for $S$ at a required scale.
In \cite{rad1} this kind of study has been done extensively 
for the similar models in the different context, and they found that
the symmetry breaking scale required here could be successfully
realized. 
We also have not sufficiently discussed the relation between this
scenario and the inflation of the universe as the next step problem.
It is necessary to embed this scenario into a suitable inflation model. 
We may construct such a kind of inflation model in the direction of
\cite{laz,infl}, in which the similar models are discussed. 
These subjects will be discussed in other places. 

\vspace*{5mm}
\noindent
This work is supported in part by a Grant-in-Aid for Scientific 
Research (C) from Japan Society for Promotion of Science
(No.~14540251) and also by a Grant-in-Aid for Scientific 
Research on Priority Areas (A) from The Ministry of Education, Science,
Sports and Culture (No.~14039205).

\vspace*{1cm}
\newpage
\noindent
{\Large\bf Appendix A}

The nonrenormalizable term in the superpotential $W_1$ can be controlled
by introducing a suitable symmetry.
One way is to impose a discrete symmetry $Z_N$ and assign its unit charge
to $S$. Then the second term in $W_1$ can be obtained as 
the lowest order term. 
We can also obtain other terms in $W_{\rm MSSM}+W_1$ by assigning the 
discrete charge to the MSSM contents suitably.
This construction is assumed in the text.
 
It may be extended by introducing one more singlet chiral superfield $\bar S$
and considering the discrete symmetry $Z_{N^\prime}~(N^\prime>N)$. 
We assign them the discrete integer charge as $q(>0)$ for $S$ and 
$-p(<0)$ for $\bar S$. If the least common multiple for $p$ and $q$
is assumed to be $pq$ and $p+q=N$, the lowest order $Z_{N^\prime}$ invariant 
term is ${d\over M_{\rm pl}^{N-3}}S^p\bar S^q$. 
Considering a $D$-flat direction: 
$\langle S\rangle=ue^{i\theta_S},~\langle\bar S\rangle
=ue^{i\theta_{\bar S}}$ and putting $\theta=p\theta_S+q\theta_{\bar S}$, 
then the feature of this direction is similar to that discussed in
the text. However, in this case we needs to introduce a coupling of $\bar S$ 
to some chiral superfields $f_\alpha$ which makes the decay of the $\bar S$ 
condensate possible. As far as the MSSM fields have no interaction 
with $f_\alpha$, the same results obtained in the text will be realized.  
The decay products $f_\alpha$ might explain the cold dark matter abundance 
if they have no other interaction which breaks the $X$ symmetry.

Another way to present $W_1$ is to assume the existence of an 
additional U(1) gauge symmetry to the MSSM gauge structure. 
We assign its integer charge with the opposite sign such as $q(>0)$ 
and $-p(<0)$ to $S$ and $\bar S$ where $p$ and $q$ is assumed to 
satisfy the condition that their least common multiple
is $pq$ and $p+q=N$. Then 
the same order term ${d\over M_{\rm pl}^{N-3}}S^p\bar S^q$ 
to the second one in $W_1$ can be obtained as the lowest order 
term constructed from $S$ and $\bar S$. 
The introduction of such a symmetry is also favored from a
view point to escape the tadpole problem and the domain wall problem,
which are expected to appear often in the models extended only by the
singlet chiral superfield \cite{rad}. Moreover,
since the extra U(1) gauge interaction imposes the $D$-flatness
$q|\langle S\rangle|=p|\langle\bar S\rangle|$, both condensate $S$ and
$\bar S$ can decay through the coupling $\lambda SH_1H_2$.
In this case, however, the introduction of the new fields is 
required from the gauge anomaly cancellation. Since these fields 
contribute to the thermal equilibrium conditions for the
SU(2) and SU(3) sphaleron interactions {\it etc.}, the discussion presented 
in section 3 cannot be directly applied.
Although it may be possible to introduce the new fields without 
changing the results obtained in the text qualitatively, 
the model will be rather complicated.

\vspace*{5mm}
\noindent
{\Large\bf Appendix B}

In this appendix we briefly explain the small neutrino mass generation 
in the $R$-parity violating models.
In the MSSM extended by $W_3$, the first term in $W_1$ or the $\mu$-term 
collaborates with $W_3$ to induce a mixing mass matrix between neutrinos 
and neutralinos in such a way as 
\begin{eqnarray}
&&{\cal M}=\left(\begin{array}{cc}
0 & M_m \\  M_m  & M_{\cal N} \\
\end{array}\right), \quad M_m=\left(
\begin{array}{cccc}
\sqrt 2g_2\langle\tilde\nu_e\rangle &\sqrt 2g_1\langle\tilde\nu_e\rangle 
& 0 & \epsilon_e \\
\sqrt 2g_2\langle\tilde\nu_\mu\rangle 
&\sqrt 2g_1\langle\tilde\nu_\mu\rangle & 0 & \epsilon_\mu \\
\sqrt 2g_2\langle\tilde\nu_\tau\rangle 
&\sqrt 2g_1\langle\tilde\nu_\tau\rangle & 0 &\epsilon_\tau \\
\end{array}\right), \nonumber  \\
&&M_{\cal N}=\left(\begin{array}{cccc}
M_2 & 0 & {1\over\sqrt 2}g_2v_1 & -{1\over\sqrt 2}g_2v_2\\ 
0  & M_1 & -{1\over\sqrt 2}g_1v_1 & {1\over\sqrt 2}g_1v_2 \\
{1\over\sqrt 2}g_2v_1 &-{1\over\sqrt 2}g_1v_1  & 0 & \mu \\ 
-{1\over\sqrt 2}g_2v_2 &{1\over\sqrt 2}g_1v_2  & \mu & 0 \\
\end{array}\right),
\label{eqf}
\end{eqnarray}
where we write ${\cal M}$ by using the basis 
$(L_\alpha,~-i\lambda_2^3,~-i\lambda_1,~\tilde H_1,~\tilde H_2)$. 
$M_{\cal N}$ corresponds to the usual neutralino mass matrix.
Nonzero sneutrino VEVs $\langle\tilde\nu_\alpha\rangle$ are
expected to be derived by minimizing the scalar potential which 
contains the soft SUSY breaking terms 
$B_{\epsilon_\alpha}\epsilon_\alpha L_\alpha H_2$ corresponding to $W_3$.
It can be estimated as $\langle\tilde\nu_\alpha\rangle\sim O(\epsilon_\alpha)$.
If we assume that both the absolute values of 
$\epsilon_\alpha$ and $\langle\tilde\nu_\alpha\rangle$ 
are much smaller than $\mu$ and $ M_{1,2}$ which 
are considered to be the order of weak scale $M_W$,
the small neutrino masses are generated by the weak scale seesaw 
mechanism \cite{rparity,hdprv}.

It is easily checked that ${\cal M}$ has two zero and 
five nonzero eigenvalues. The four nonzero eigenvalues correspond to
those of the neutralinos and they are $O(M_W)$. 
On the other hand, as discussed in \cite{hdprv}, 
the tree-level mass eigenvalues are characterized 
by the quantities
$\Lambda_\alpha\equiv\epsilon_\alpha\langle 
H_1\rangle+\mu\langle\tilde\nu_\alpha\rangle$
in the effective neutrino mass matrix. 
Using this $\Lambda_\alpha$, the smallest one of the nonzero eigenvalues is
written as $(M_1g_1^2+M_2g_2^2)\vert \vec{\Lambda}\vert^2/4{\rm
det}(M_{\cal N})$. The remaining massless states are known to 
become massive by taking account of the radiative corrections.
If we impose the constraints on the mass and mixing 
required for the explanation of the neutrino oscillation data \cite{sol,atm},
it is shown that $\epsilon$ should satisfy the 
condition such as $\vert\epsilon/\mu\vert
\sim O(10^{-4\sim -3})$ by using the numerical analysis \cite{hdprv}.

The model with the bilinear $R$-parity violating term can be extended by
introducing the generation dependent extra U(1) gauge symmetry at the
TeV region \cite{rparity1}. In this case, the neutrino mass degeneracy
can be resolved at the tree-level and then the mass and mixing of
neutrinos can be directly related to $\epsilon$.
The largest mass eigenvalue can be written as 
$O(\vert\epsilon^2/\mu\vert)$,
where the gaugino masses are assumed to be $O(\vert\mu\vert)$.
By imposing the condition for the $\nu_\tau$ mass required from the
explanation of the atmospheric neutrino anomaly, we can obtain a
condition $\vert\epsilon/\mu\vert \sim O(10^{-6})$ for a reasonable
value of $\mu$. The value of $\vert\epsilon\vert$ can be smaller than 
the one in the previous example by two or three order of magnitude.

\newpage

\end{document}